# Control Over Fano Parameter in Grating and One-Dimensional Photonic Crystal Cavity

*Pratip Ghosh[1] and Akshay K. Naik[1*]*

[1]*Centre for Nano Science and Engineering,*

*Indian Institute of Science, Bengaluru-560012, Karnataka, India*

*\*Corresponding author: anaik@iisc.ac.in*

**Abstract**.

Fano resonances are sharp asymmetrical spectral peaks which are now ubiquitous in nanophotonics. The high sensitivity of these resonances to system parameter has been exploited to improve light matter interaction and in applications such as sensing, filters and on-chip processing. The ability to dynamically change the Fano slope and spectral phase would enable optimization of the device parameters post fabrication for various applications. Here we demonstrate such a control over the Fano resonance in a one-dimensional photonics crystal cavity integrated on a silicon waveguide -grating platform. In our device, Fano resonance arises due to interference between cavity mode and an oscillatory background due to grating coupler. The dynamics tuning of Fano asymmetric parameter is achieved using thermos-optic effect in silicon. We experimentally tune the Fano parameter from ~-3.2 to +1.7 achieving a highest extinction ratio of 21.6 dB and spectral slope of 108dB/nm. All the above is achieved in an ultra-compact design with simple fabrication and with multiple cavities or feedback elements. The steep slope offers distinct advantage over conventional cavity for sensing and modulation applications and the tunability enables dynamic control over gain, dynamic range, bandwidth and noise coupling.

**Introduction:** The interference between discrete localized state and a continuum of state results in distinctively asymmetrical peaks which are known as Fano resonances[1–3]. Since the original observation, these peaks have now been observed to wide variety of systems including photonic devices[3–7]. In optics, these effects are observed when a narrow cavity mode overlaps with a broad waveguide mode [1,2]. Unlike the typical symmetrical Lorentzian resonance peaks, Fano resonances exhibit rapid transition between total transmission and total reflection [8]. This high contrast change with wavelength is useful for many applications including modulation[9,10] and sensing applications[10–13]. These resonances enable higher sensitivity without bandwidth loss associated with high Q cavities and reduced chirp in intensity modulation that are very important for high-speed communication[14].

There are multiple approaches to engineer these Fano resonance in nanophotonic devices. For example, metasurfaces and metamaterials composed of coupled resonant elements can yield narrow Fano features through collective interference[3]. Similarly, high-Q microresonator (such as a disk or ring)[11] side-coupled to a bus waveguide[15–18] create interference between the resonator mode and the waveguide continuum. Such coupled-resonator schemes have been used to demonstrate compact thermo-optic Fano switches[19–21]. A defect cavity in a 1D Photonic crystal (PhC) waveguide produces an interaction between the localized cavity mode and the propagating waveguide mode, giving rise to an asymmetric transmission spectrum[6,14,20,22,23].

However, ability to tune the Fano parameter remains an ongoing endeavor. Such a control post fabrication would enable tradeoffs between sensitivity and dynamic range of sensors, optimize bit error rate in high-speed modulation and more reconfigurable systems for neuromorphic photonics[24].

Here, we demonstrate Fano resonances in a one-dimensional photonic crystal nanocavity etched directly into a silicon-on-insulator (SOI) waveguide. In contrast to conventional designs[4,7,9,18,25], the Fano interference in our device arises naturally from the combined effect of waveguide-cavity mismatch and grating-coupler-induced reflections. These introduce a broadband oscillatory background without the need for additional resonators[4,7] or interferometric structures. This compact configuration enables strong Fano interference within an ultra-small footprint while maintaining fabrication simplicity.

We demonstrate controlled tuning of the cavity resonance and modulation of the Fano lineshape, by exploiting the thermo-optic effect of silicon[26–28]. The resulting change in the relative phase of the interfering pathways allows continuous tuning of the Fano asymmetry parameter[29]. As the device temperature increases, the spectrum changes from a nearly symmetric, Lorentzian-like response to a strongly asymmetric Fano profile. We also demonstrate that the Fano parameter can be modified independently of the resonance wavelength by adjusting the fiber position at the grating coupler. This provides the interference control without shifting the cavity resonance wavelength. Unlike previous reports, that primarily emphasize extinction or switching performance, we focus on explicitly characterizing and controlling the Fano parameter within a compact photonic crystal platform. These results establish a simple approach to realize tunable Fano resonances in integrated photonic circuits.

**Device Fabrication:** We implement the device in photonic crystal (PhCs) in the waveguide. PhCs are structures with a periodic variation in the dielectric function, designed to manipulate the light propagation. This periodicity creates an optical bandgap at the two ends. A defect in the middle of this periodic structure gives rise to a localized optical mode within the bandgap. The mirror at the two ends of the photonic crystals can be designed to reduce the reflectivity/transmission as required for the Fano resonance. In our work, the unit cell consists of a rectangular silicon region with a centrally located elliptical air hole. The periodicity of these unit cells is 460 nm. A defect is introduced by altering the major axis of the central ellipse, which traps light within the cavity.

The fabrication of the Fano-resonant photonic device was carried out on a silicon-on-insulator (SOI) substrate comprising a 220 nm thick silicon device layer on top of a 2 μm buried oxide layer. A three-step electron-beam lithography (EBL) process was used to define the optical and thermal components of the device. In the first step, a negative Ma-N resist was used to pattern the photonic crystal cavity, input/output waveguides, and grating coupler regions. The developed pattern was transferred onto the silicon layer using reactive ion etching (RIE). The second lithography step, using PMMA as the resist, was used to define the grating couplers. A shallow etch was performed using RIE to optimize coupling efficiency around 1560 nm. In the final lithography step, double layer of PMMA was used to define the layout for metal microheaters near the cavity region. Cr/Au microheaters, with a total thickness of 100 nm, were deposited using thermal evaporation, followed by liftoff. These heaters were placed approximately 2 μm away from the cavity to enable localized thermal tuning of the resonance. This fabrication sequence ensures precise control over both the optical transmission characteristics and the tunability of the Fano resonance. Figures 1(a–d) show the fabricated one-dimensional photonic crystal cavity, waveguide, and grating couplers.

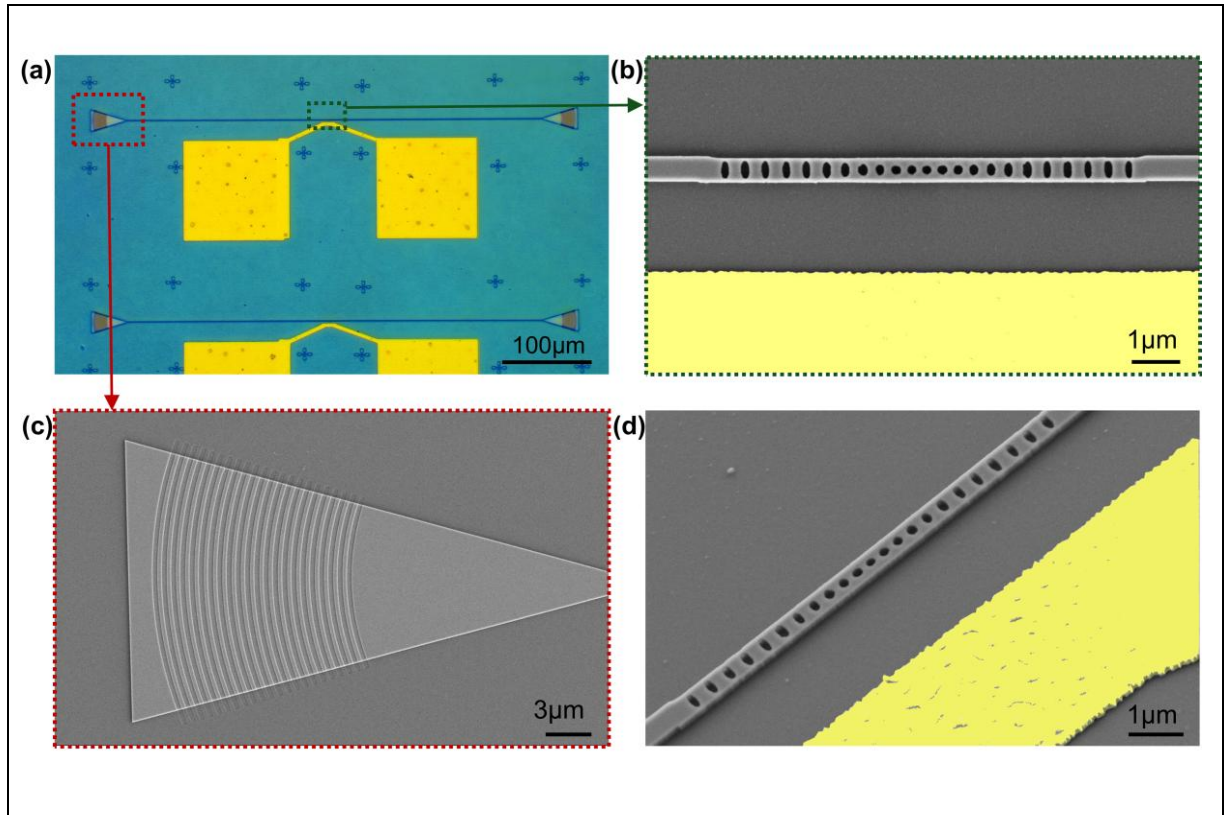

*Figure 1: (a). optical image of the fabricated devices with the heating pads, (b) SEM image of the zoomed in image of the cavity region and (c) grating coupler, (d) Bird's eye view of the cavity and heater region.*

The waveguide, 530 nm wide and 400 μm long, is intentionally mismatched with the cavity to ensure weak coupling, which contributes to the formation of a continuum of states. The grating couplers used in the design not only serve as coupling elements but also as strong reflectors, giving rise to a weak Fabry-Pérot cavity within the waveguide. This reflection introduces an oscillatory background in the transmission spectrum. The combined effect of the waveguide-cavity mismatch and the grating-induced Fabry-Pérot resonance establishes the non-resonant continuum pathway.

Interference between this continuum and the discrete resonance of the cavity gives rise to the Fano resonance. Here, the weight factors corresponding to the resonant and non-resonant optical paths govern the nature and strength of the interference, playing a crucial role in shaping the Fano lineshape observed in our system.

**Experimental Result:** A tunable laser source is used to sweep the input wavelength over a range of 1510–1630 nm. The laser output is connected via optical fiber to a fiber polarization controller (FPC) to ensure optimal polarization for coupling. Light is coupled into and out of the photonic chip using grating couplers, which are designed for peak efficiency near 1560 nm. The input fiber is positioned at an angle of 10° relative to the surface of the chip to align with the grating coupler, and similarly, the output fiber is placed at 10° angle and connected to an optical power meter to record the transmission signal. Figure 2(a). shows schematic of the experimental setup and inset shows schematic of the device.

The transmission spectrum of one of the fabricated devices exhibiting a characteristic Fano resonance is shown in Figure 2(b). The measured extinction ratio between the transmission peak and dip is approximately 21.6 dB. The wavelength separation between the maximum and minimum transmission points is about 0.21 nm, corresponding to a spectral slope of 108 dB/nm, indicative of a sharply varying response suitable for switching and sensing applications. The inset shows the transmission spectrum plotted on a linear scale, along with the corresponding Fano fit

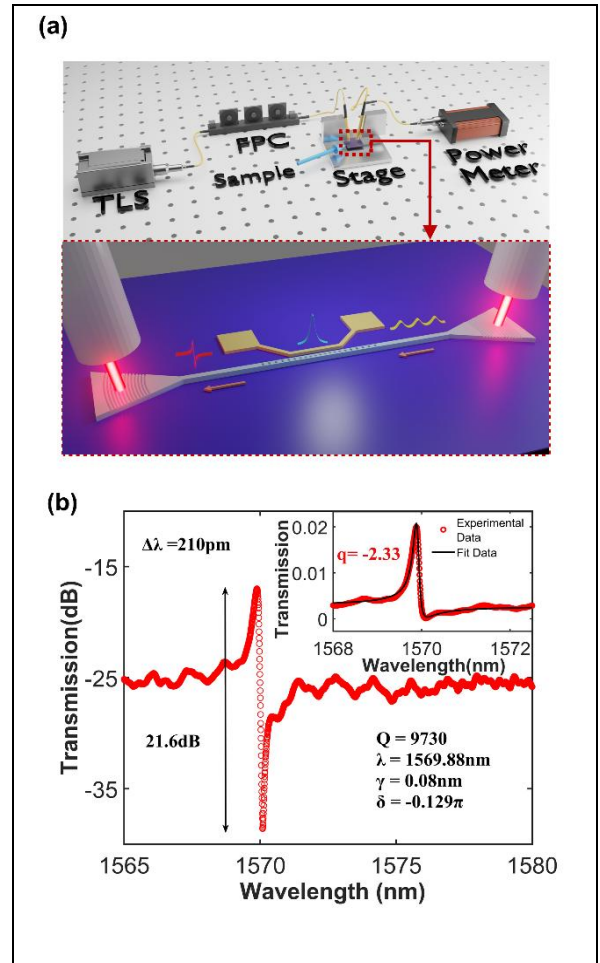

*Figure 2: (a) Schematic Experimental set-up to measure transmission spectrum. Schematic of the device(inset). (b) Transmission response of device showing Fano resonance. The fitted data (inset) shows q= -2.33.*

used to extract the Fano parameter $q = -2.33$.

**Origin of Fano resonance and Analytical model:** To investigate the origin of the Fano resonance, we used temporal coupled-mode theory (TCMT)[30,31] . The model has two important components viz.,

a) a high-quality factor optical cavity with transmission given as

$$T(\omega) = \frac{|S_2^-|^2}{|S_1^+|^2} = \frac{4\gamma_1\gamma_2}{(\omega - \omega_0)^2 + \gamma_t^2}$$
$$= \frac{4\gamma^2}{(\omega - \omega_0)^2 + \gamma_t^2} \quad (1)$$

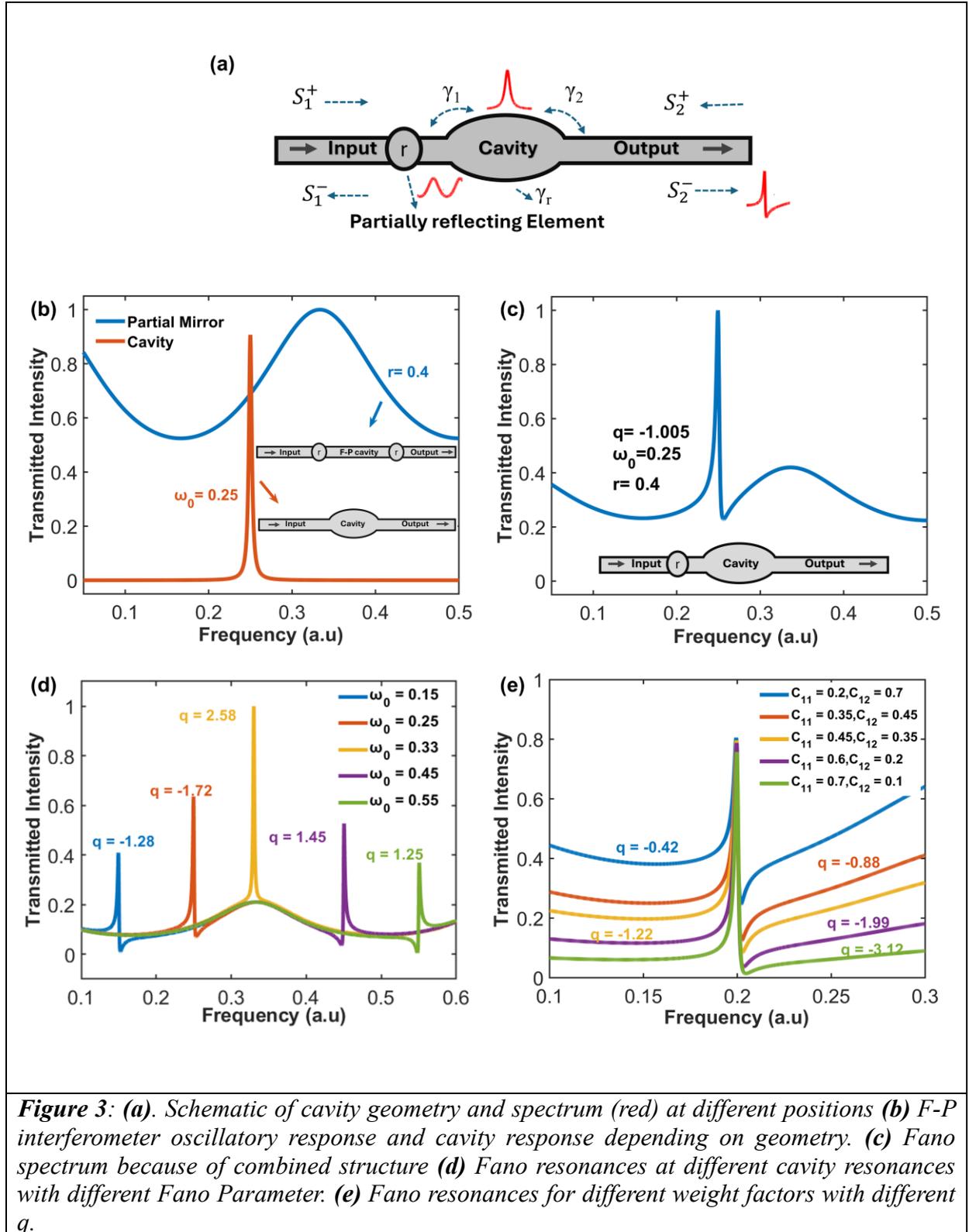

*Figure 3*: *(a)*. *Schematic of cavity geometry and spectrum (red) at different positions* *(b)* *F-P interferometer oscillatory response and cavity response depending on geometry.* *(c)* *Fano spectrum because of combined structure* *(d)* *Fano resonances at different cavity resonances with different Fano Parameter.* *(e)* *Fano resonances for different weight factors with different q.*

Where, $S_1^+$ and $S_1^-$ denote the forward and backward propagating field amplitudes at the input port, while $S_2^+$ and $S_2^-$ represent the corresponding amplitudes at the output port[32–34]. The total decay rate of the cavity mode is given by,

$$\gamma_t = \gamma_1 + \gamma_2 + \gamma_r$$

$\gamma_1$ and $\gamma_2$ are the coupling rates to the input and output waveguides, respectively, and $\gamma_r$ accounts for radiative decay from the cavity. We assume that the cavity couples only to the input and output waveguides, with equal coupling strengths ($\gamma_1 = \gamma_2 = \gamma$), and that other loss channels are negligible.

b) two partially reflecting elements around the cavity[35] [11] with transmittivity,

$$t_{FP} = \frac{(1-r^2)e^{-i\delta/2}}{r^2 e^{-i\delta} - 1} \quad (2)$$

Where, the reflectivity of both mirrors is $r$, and the separation between them is $l$. $\delta/2$ denotes the phase shift accumulated as the waveguide mode propagates between the two partially reflecting elements. The transmission spectra when the two components are present individually are show in figure 3(b). The cavity produces a sharp Lorentzian response while the two partially reflecting mirrors form a Fabry-Perot cavity with a broad resonance. Combining the two elements yields a sharp resonance on top of a broad background peak as required for a Fano resonance (see figure 3c). The response is governed by the equation (see supplementary information for details)

$$t(\omega) = \sqrt{C_{11}} \frac{2\sqrt{\gamma_1 \gamma_2}}{i(\omega - \omega_0) + \gamma_t} + \sqrt{C_{12}} \frac{(1-r^2)e^{-i\delta/2}}{r^2 e^{-i\delta} - 1} \quad (3)$$

where, the coefficients $C_{11}$ and $C_{12}$ represent the relative contributions of the two interfering optical pathways. Here, $C_{11}$ represents the amplitude coefficient for the resonant pathway, while $C_{12}$ corresponds to the non-resonant optical path amplitude coefficient. By manipulating the relative phase of the response of the two resonances, the Fano parameter can be manipulated (figure 3(d)). As shown in Figure 3(e), when the resonant path dominates ($C_{11} = 0.7, C_{12} = 0.1$), the Fano parameter is small, resulting in a Lorentzian-like symmetric peak. Conversely, when the non-resonant (direct) becomes dominant ($C_{11} = 0.2, C_{12} = 0.7$), the Fano parameter becomes large, leading to a weaker interference effect. The most prominent Fano interference can be observed in the graph when the Fano parameter is close to ±1, though not exactly equal to ±1. In this case, for the condition $C_{11} = 0.45, C_{12} = 0.35$. These two weight factors thus play a crucial role in determining the shape and strength of the Fano resonance.

**Thermo-optic effect and tuning of Fano resonance:** To demonstrate dynamic control of the Fano lineshape, we applied a DC bias to the gold micro-heater positioned near the optical cavity. This induced localized heating and altered the effective refractive index of the silicon cavity via the thermo-optic effect. The transmission spectra were recorded for heater power range from 0 mW to ~132 mW. The temperature-induced change in the refractive index was modelled using the relation,

$$n(T) = n_0 + \frac{dn}{dT}(T - T_0) \quad (4)$$

where, $n_0$ is the refractive index at room temperature ($T_0$). We assume a thermo-optic coefficient of silicon, (TOC) $\frac{dn}{dT} = 1.893 \times 10^{-4}/°C$. As the temperature increases due to electrical heating, the effective refractive index of the cavity material increases, leading to a redshift of the resonance wavelength. This behaviour is consistent with the positive thermo-optic coefficient of silicon, as described by Eq. (4). This behaviour is illustrated in Fig. 4., where the Fano resonance evolves significantly with increasing power.

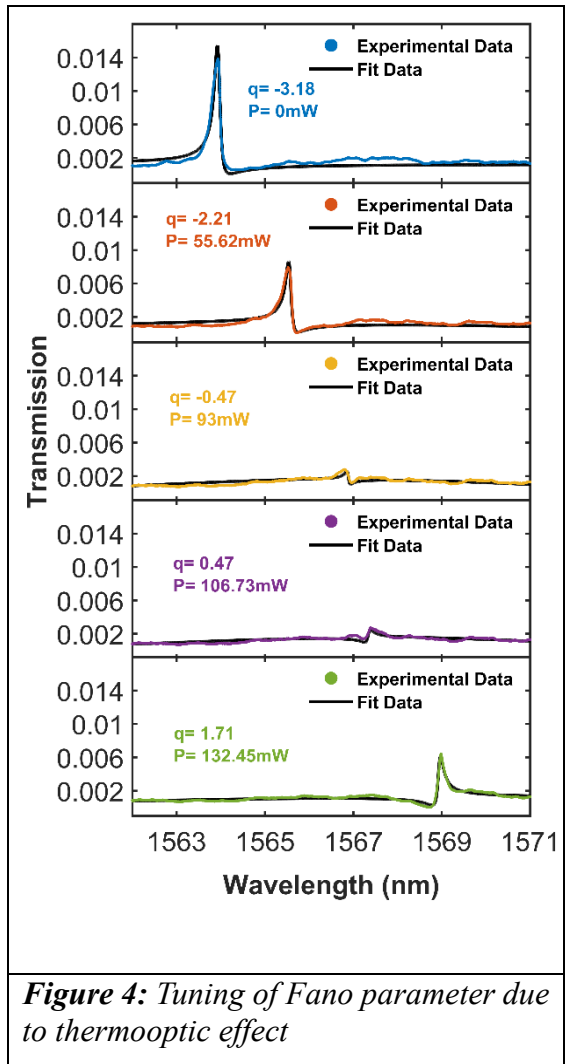

*Figure 4: Tuning of Fano parameter due to thermooptic effect*

At zero nominal heating, the transmission spectrum exhibits a Fano parameter $q = -3.18$, which corresponds to an almost symmetric, Lorentzian-like peak. As the electrical power is gradually increased, the Fano lineshape becomes increasingly asymmetric, and the parameter q is continuously tuned, eventually reaching a value of $q = +1.71$. The complete set of transmission spectra recorded at intermediate heating powers, along with the corresponding Fano fits and extracted parameters, are shown in the Supplementary Information.

At intermediate heating (93mW and 107mW), the Fano parameter approaches ±1, indicating the regime of strongest interference. This demonstrates that both the spectral asymmetry and extinction ratio of the Fano resonance can be effectively modulated by electrical control of the local temperature.

To further quantify the thermo-optic tuning behaviour, we extract key performance metrics as a function of the applied electrical power. Figure 5(a) shows the resonance wavelength shift as a function of heating power exhibiting a clear monotonic redshift. This is consistent with the positive thermo-optic coefficient of silicon, and the localized temperature increase near the cavity. A linear fit to the experimental data yields a tuning efficiency of 0.0347 nm/mW. The modest tuning efficiency arises from the ~2 μm separation between the heater and the cavity, leading to reduced thermal coupling. During thermo-optic tuning, the intrinsic quality factor and corresponding cavity decay rate remain nearly constant over most of the applied power range. Minor fluctuations observed at certain bias conditions (Figure 5b) are attributed to fitting uncertainties arising from the asymmetric Fano lineshape and variations in signal-to-noise ratio, rather than any significant change in the intrinsic cavity losses. Figure 5(c) presents the variation of extinction ratio with electrical power, demonstrating that the resonance depth can be selectively enhanced at specific bias conditions, where the transmitted signal contrast is maximized. Such operating points are particularly useful for applications requiring high signal visibility. In addition, Fig. 5(c) shows the evolution of the spectral slope, defined here as the ratio of extinction ratio to the wavelength separation between the maximum and minimum transmission points. Due to the asymmetric nature of the Fano lineshape, the slope can assume both positive and negative values depending on the relative positions of these extrema. Figure 5(d) shows the variation of the Fano asymmetry parameter $q$ with heater power, highlighting the continuous tuning of the interferenc condition achieved through electrical control. The corresponding evolution of the relative phase $\delta$, obtained from the relation $q = \cot\delta$, indicates a continuous transition of the phase from negative to positive values with increasing heater power. The ability to electrically tune the extinction ratio, spectral slope, and Fano parameter enables the selection of optimal operating conditions, where high extinction or steep spectral response can be exploited for sensitive measurements and efficient modulation. Together, these results demonstrate that controlled electrical tuning provides a practical means to optimize device performance for application-specific requirements.

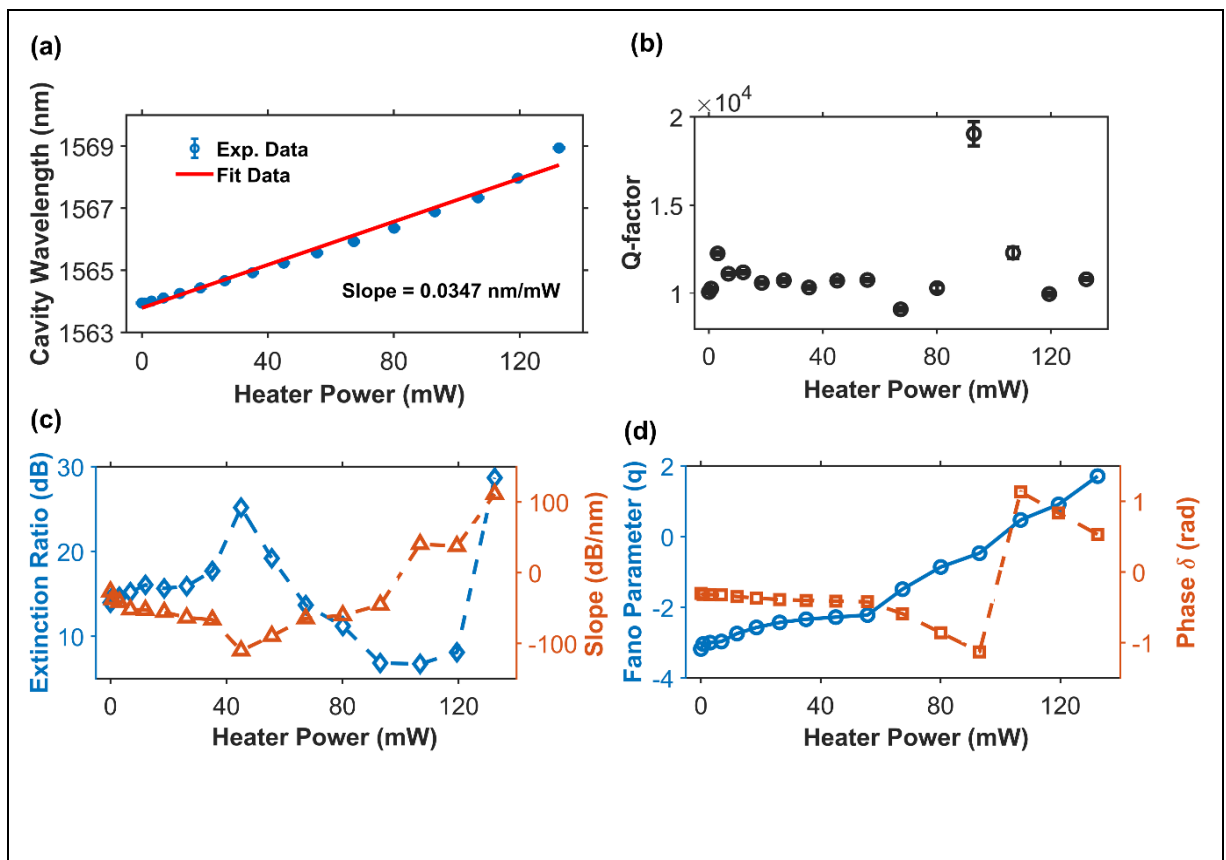

*Figure 5: (a) Heater Power vs Cavity resonance wavelength with linear fitting data. Heater Power vs (b) optical Q-factor. (c) variation of extinction ratio and slope of the fano spectrum. (d) change in the Fano parameter and relative phase.*

Furthermore, the tuning process is reversible i.e. when the applied voltage is removed and the device returns to ambient temperature, the resonance peak and Fano lineshape recover their original form. This indicates that the device operates without observable hysteresis over the tested power range, making it suitable for repeated cycling. Such dynamic and reversible control of the Fano lineshape highlights the robustness of the platform and its suitability for integrated photonic applications. The ability to modulate resonance properties channel provides a direct route to tuning the Fano parameter without perturbing the intrinsic cavity resonance. Experimentally, this can be achieved by adjusting the excitation or collection geometry, such as changing the relative position or incident angle of the input fiber with respect to the grating coupler. Both modifications with relatively low electrical power makes this device attractive for optical switching, filtering, and modulation. The demonstrated control over the Fano parameter, spanning regimes of destructive to constructive interference, provides an additional degree of freedom for tailoring device response, which is particularly beneficial for on-chip signal processing, programmable photonic networks, and neuromorphic photonics.

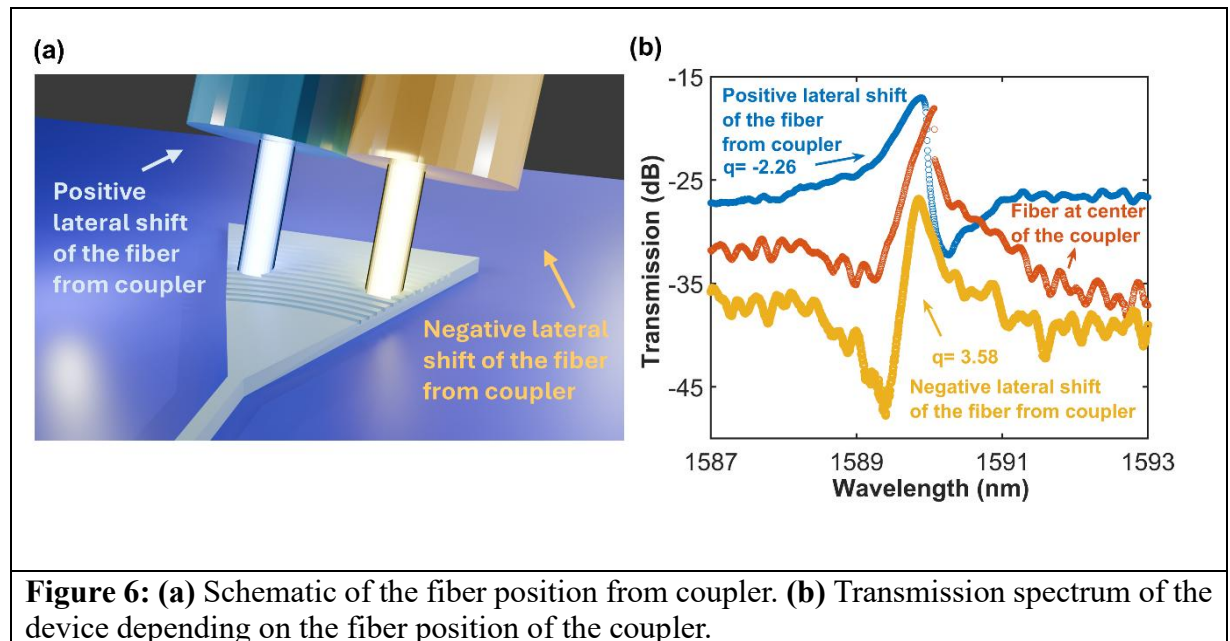

**Figure 6: (a)** Schematic of the fiber position from coupler. **(b)** Transmission spectrum of the device depending on the fiber position of the coupler.

Beyond thermo-optic tuning, the Fano asymmetry parameter can also be controlled independently of the resonance wavelength by modifying the relative strength of the interfering optical pathways. Since the Fano lineshape arises from interference between the resonant response and a non-resonant background, tailoring the coupling conditions of the continuum selectively alter the oscillatory background contribution while leaving the cavity resonance wavelength essentially unchanged. From an on-chip perspective, similar control can be realized by engineering the reflectivity of the grating couplers, introducing controlled loss or phase modulation in the access waveguide, or varying the relative coupling strengths at the input and output ports.

Figure 6(a). shows the schematic of the experimental condition of the fiber at different position relative to the coupler (blue and yellow two different positions). Figure 6(b). experimentally demonstrates this concept by showing the transmission spectra measured for two different positions of the input fiber relative to the grating coupler. A clear modification of the oscillatory background is observed when

| Reference (System) | Tuning Mechanism | q range | ER (dB) | Slope (dB/nm) | Notes/Complexity |
|---|---|---|---|---|---|
| Zhang et al., 2016 (Si microring + FP cavity)[17] | All-optical (TPA-induced nonlinear thermal-optic effect) | - | 22.54 | 250.4 | Strong slope/ER, but requires feedback FP cavity structure |
| Zhao et al., 2016 (Si ring+feedback)[7] | Thermo-electric | - | 30.8 | 226.5 | High ER (30.8 dB) achieved; design uses feedback loop |
| Zhang & Yao, 2018 (Si microdisk+MZI)[19] | Thermal (microheater) | -0.98 to 0.95 | 22.8 | ~41 | $q$ crosses zero; compact but requires MZI coupling |
| Gu et al., 2020 (Si ring+air-hole)[16] | Geometry/Fabrication (air-hole size change) | 1 to ∞ (only positive) | >20 | >400 | Very steep (400+ dB/nm), uses engineered airhole fabrication complexcity |
| Mehta 2013 (Si PhC, $q$ fixed by fiber)[36] | fiber coupling |  | Up to ~17 |  | Improved Fano contrasts, but no dynamic q control |
| **This work** | Thermo-optic + fiber coupling | -3.18 to +1.71 | 21.6 d | 108 | Compact design; decouples phase vs. continuum tuning |

**Table 1.** Comparison of representative Fano resonance implementations in integrated photonic platforms, including system type, tuning mechanism, Fano parameter control, extinction ratio, and spectral slope.

the fiber position is changed, resulting in a pronounced change in the extracted Fano asymmetry parameter from $q = 3.58$ to $q = -2.26$. Importantly, the resonance wavelength remains essentially unchanged in both cases, confirming that the cavity mode itself is not perturbed. The observed sign reversal of $q$ corresponds to a complete change in the interference condition between the resonant and non-resonant pathways, arising solely from the modified background contribution. To further support this interpretation, an analytical model based on temporal coupled-mode theory is provided in the Supplementary Information, reproducing the same qualitative evolution of the Fano lineshape under variations of the background pathway. This confirms that the observed tuning of the Fano parameter is governed by interference control rather than intrinsic cavity tuning.

Table 1 summarizes the performance of the proposed device in comparison with representative Fano resonance platforms. The demonstrated system combines a compact and fabrication-friendly geometry with controlled tuning of the Fano asymmetry parameter $q$, achieved through thermo-optic modulation and independent control of the background contribution. The device further exhibits a high extinction ratio and a steep spectral slope, along with tunability of the operating wavelength. This combination of features, realized within a single-cavity architecture, provides a flexible platform for optimizing spectral response in integrated photonic applications.

**Conclusion:** In summary, we experimentally demonstrate dynamic control of a Fano resonance in a one-dimensional photonic crystal nanocavity integrated directly onto a silicon waveguide. Unlike many previously reported approaches that rely on deliberately engineered multi-resonator systems or external interferometric coupling schemes, the Fano resonance in our device emerges naturally from the combined effect of waveguide–cavity mismatch and grating-coupler-induced reflections. These reflections introduce a broadband oscillatory background that interferes with the cavity resonance, enabling strong Fano interference within a highly compact footprint.

By exploiting the thermo-optic effect in silicon, the Fano asymmetry parameter is continuously tuned from $q = -3.18$ to $q = +1.71$, accompanied by a controlled redshift of the cavity resonance. The grating couplers serve a dual function as efficient input–output interfaces and partially reflecting elements, eliminating the need for additional mirrors or complex coupling architectures. This intrinsic formation of the non-resonant background allows effective control over the relative weights of the resonant and non-resonant pathways using a simple device geometry.

The fabricated device exhibits a high extinction ratio of 21.6 dB and a steep spectral slope of 108 dB/nm, highlighting the sharp spectral features enabled by this interference mechanism. Due to its simplified geometry, ultra-small footprint, and reliance on standard fabrication processes, the proposed platform offers a scalable and fabrication-tolerant route toward tunable Fano-based functionalities. These attributes make it well suited for compact optical modulators, switches, and signal-processing elements in integrated photonic circuits, where design simplicity and footprint are critical.

**Acknowledgment:**

The authors thank Prof. Shankar K. Selvaraja for his insightful comments and discussions. The authors acknowledge the support of the Department of Science and Technology (DST), Government of India, under the National Quantum Mission (NQM)

# Supporting Information for:

# Control Over Fano Parameter in Grating and One-Dimensional Photonic Crystal Cavity

*Pratip Ghosh & Akshay K. Naik*

*Centre for Nano Science and Engineering, Indian Institute of Science*

**Fabrication Process Flow:** The devices were fabricated on a commercially available SOI substrate with a 220 nm silicon device layer. The overall process is CMOS-compatible and relatively straightforward. The sample was first cleaned using a Piranha solution, followed by spin-coating a negative e-beam resist (MaN-2401) with a thickness of 110 nm. Electron-beam lithography was then performed to define the device patterns. After development, resist remained only in the device regions, leaving most of the silicon surface exposed.

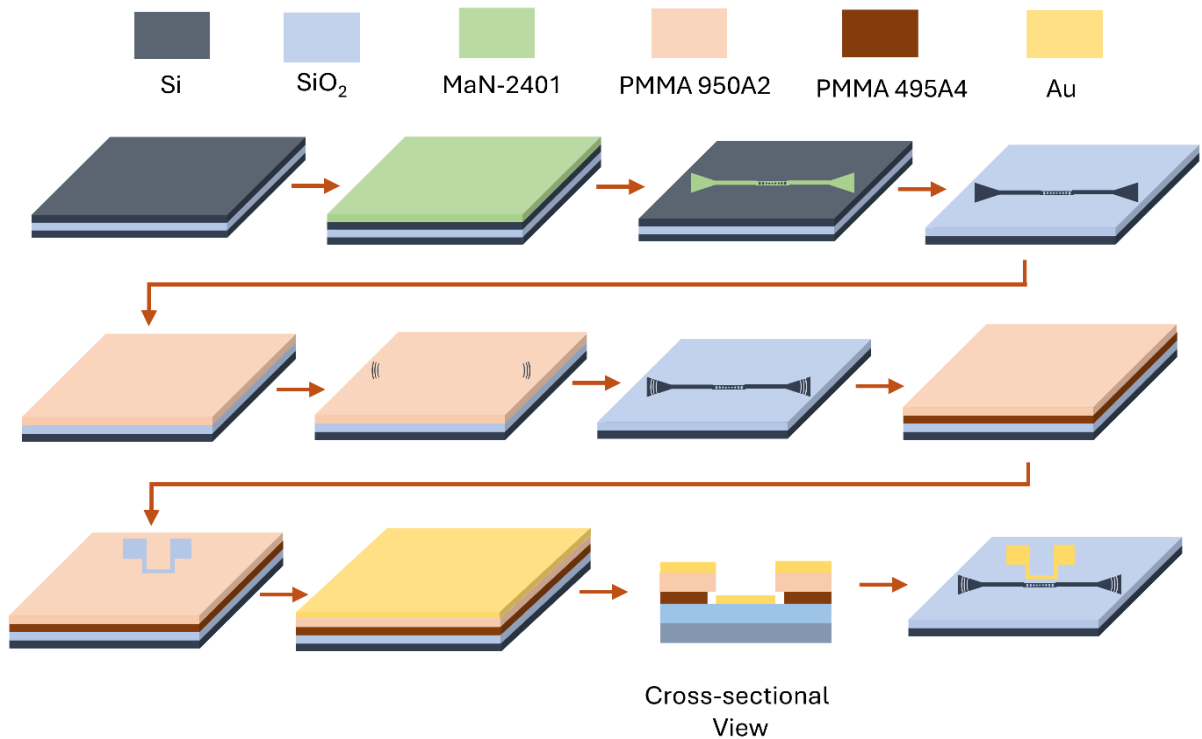

**Fig.S1.** Fabrication Process Flow for the Device.

Using fluorine-based chemistry in a reactive ion etching (RIE) system, the exposed 220 nm silicon layer was etched. Subsequently, the sample underwent oxygen ashing and another Piranha cleaning step. For the second lithography step to define the grating couplers, a positive resist (PMMA 950A2) was spin coated. After development, the silicon in the coupler region was partially etched by 70 nm to enable efficient coupling of light around 1560 nm.

To define the microheaters, a bilayer PMMA resist process was employed, consisting of PMMA 495A4 followed by PMMA 950A2 to create an undercut profile that improves the lift-off process. After electron-beam lithography for the heater patterns, 100 nm of gold (Au) was deposited using e-beam evaporation. The complete fabrication flow is illustrated in Figure S1.

**FEM Simulation:** To better understand the role of geometry in the observed spectral response, we performed Finite Element Method (FEM) simulations using COMSOL Multiphysics. The simulated structures replicate the fabricated device geometry. In this configuration (Fig. S2), we modelled an ideal photonic crystal cavity connected to input and output waveguides. The corresponding transmission spectrum, shown in Fig. S2, exhibits a symmetric Lorentzian resonance peak, characteristic of a high-Q cavity bounded by highly reflective mirrors arising from the photonic crystal bandgap. Ensuring that the geometry can give only the resonance frequency.

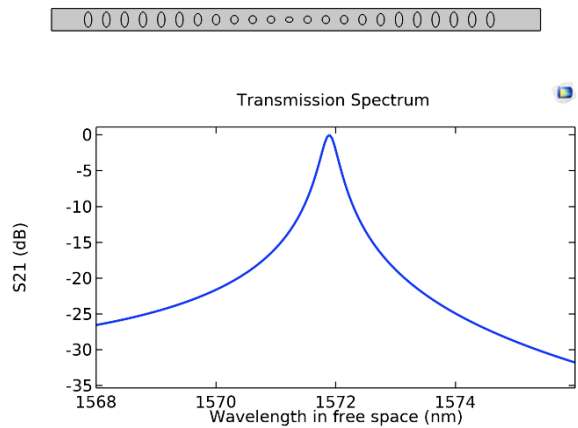

**Fig.S2.** FEM Geometry and simulation of a normal 1D Photonic Crystal Cavity

To support the experimental observations, we performed an analytical estimation of the temperature-induced refractive index change based on the measured resonance shift. The refractive index variation with temperature is described by,

$$n(T) = n_0 + \frac{dn}{dT}(T - T_0) \qquad (1)$$

Where, $n_0$ is the refractive index at room temperature ($T_0 = 300K$). We have considered the thermo-optic co-efficient (TOC) of silicon $\frac{dn}{dT} = 1.893 \times 10^{-4}/°C$. From Fig. 5(a) in the main text, the resonance wavelength shifts from $\lambda_0 = 1563.946$nm at 0mW to $\lambda = 1568.938$nm at 132mW corresponding to a wavelength shift of $\Delta\lambda = 4.992$nm. To find out the change in the refractive index is estimated as,

$$\frac{\Delta \lambda}{\lambda} = \frac{\Delta n}{n} \qquad (2)$$

$$\Delta n = n \frac{\Delta \lambda}{\lambda} \qquad (3)$$

$$\Delta n = 3.48 \times \frac{4.992}{1563.946} \approx 0.0111$$

Using the thermo-optic relation, the corresponding temperature change is estimated to be $\Delta T \approx 58.67K$. This indicates that a temperature increase of approximately 58.67 K is required to achieve the observed tuning of the Fano asymmetry parameter from $q = -3.18$

**Analytical Representation:** To investigate the origin of the Fano resonance, we employed temporal coupled-mode theory (TCMT). We consider a simplified model consisting of a single-mode optical cavity coupled to an input and an output waveguide, as illustrated in Fig. S3a. Let $S_1^+$ and $S_1^-$ denote the forward and backward propagating field amplitudes at the input port, while $S_2^+$ and $S_2^-$ represent the corresponding amplitudes at the output port. The total decay rate of the cavity mode is given by,

$$\gamma_t = \gamma_1 + \gamma_2 + \gamma_r \qquad (2)$$

where $\gamma_1$ and $\gamma_2$ are the coupling rates to the input and output waveguides, respectively, and $\gamma_r$ accounts for radiative decay from the cavity. This framework provides a foundation for understanding the interference between discrete and continuum states that gives rise to the Fano lineshape. According to TCMT[30], assuming weak coupling, the equations can be written as,

$$\frac{d\alpha}{dt} = -i\omega_0 \alpha - \alpha \gamma_t + \sqrt{2\gamma_1} S_1^+ \qquad (3)$$

For input frequency $\omega$ the equation can be written as,

$$\frac{d\alpha}{dt} = -i(\omega - \omega_0)\alpha - \alpha \gamma_t + \sqrt{2\gamma_1} S_1^+ \qquad (4)$$

$$S_1^- = -S_1^+ + \sqrt{2\gamma_1}\alpha \qquad (5)$$

$$S_2^- = \sqrt{2\gamma_2}\alpha \qquad (6)$$

Where $\alpha$ is the field amplitude in the cavity and $\omega_0$ is the resonance frequency (localized mode) of the cavity. From equation (4) and equation (6), we can get the transmission spectrum,

$$T(\omega) = \frac{|S_2^-|^2}{|S_1^+|^2} = \frac{4\gamma_1 \gamma_2}{(\omega - \omega_0)^2 + \gamma_t^2} = \frac{4\gamma^2}{(\omega - \omega_0)^2 + \gamma_t^2} \qquad (7)$$

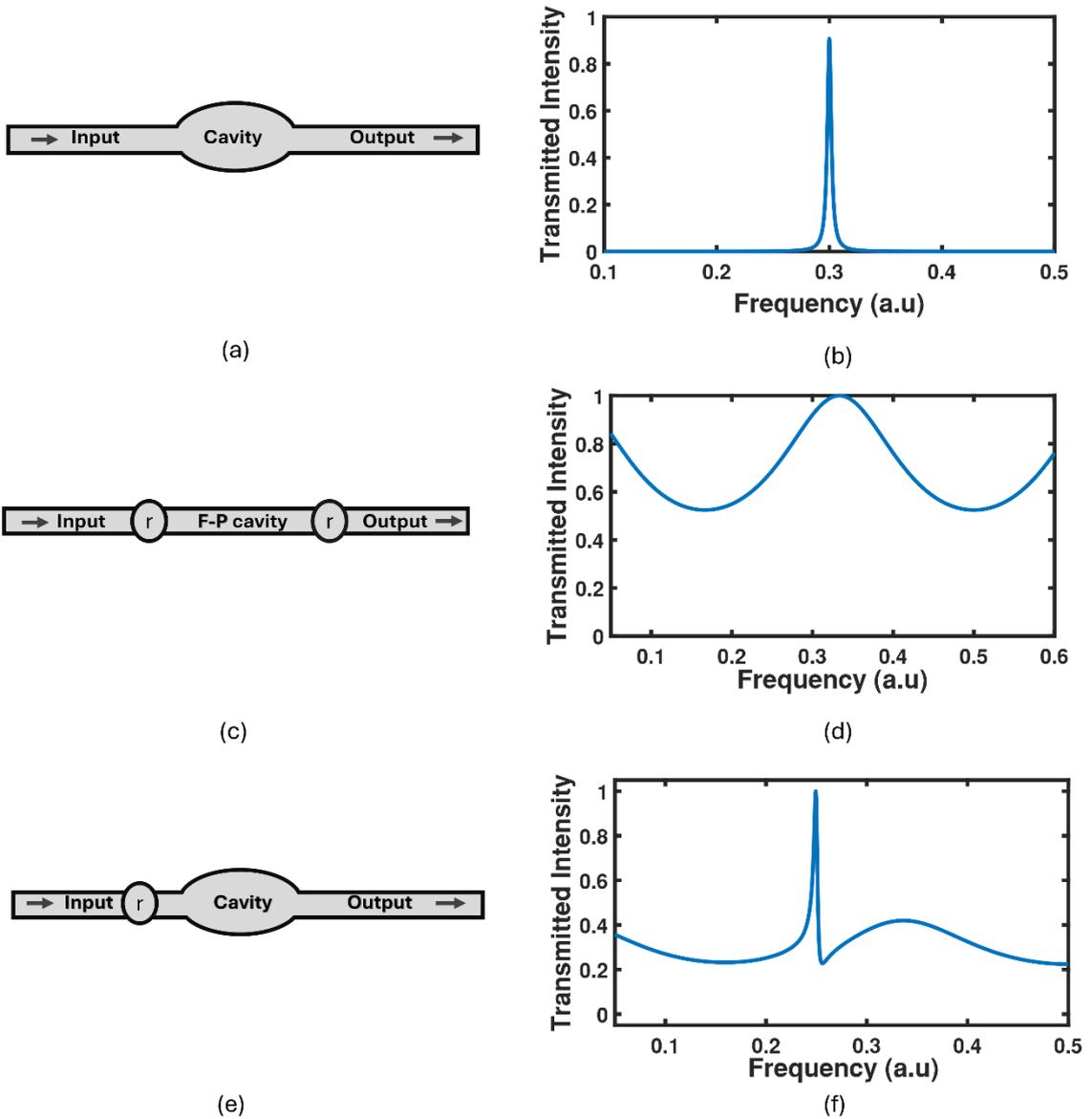

**Fig.S3**. (a), (c), (e) schematic of the geometries and (b), (d), (f) corresponding transmission response.

In the first case, we assume that the cavity couples only to the input and output waveguides, with equal coupling strengths ($\gamma_1 = \gamma_2 = \gamma$), and that other loss channels are negligible. Under these conditions, the cavity energy decays symmetrically into the waveguides. The resulting transmission spectrum at the output waveguide exhibits a Lorentzian lineshape centered at the cavity resonance frequency $\omega_0$. Equation (7) captures this behaviour, yielding a maximum transmission at $\omega = \omega_0$ (Fig.S3b.).

Next, we consider a different geometry comprising a single-mode input waveguide with two partially reflecting elements placed between the input and output ports, effectively forming a Fabry–Pérot interferometer, as depicted in Fig. S3c. Let the reflectivity of both mirrors be $r$, and the separation between them be $l$. In this configuration, the transmitted amplitude can be expressed as[35]:

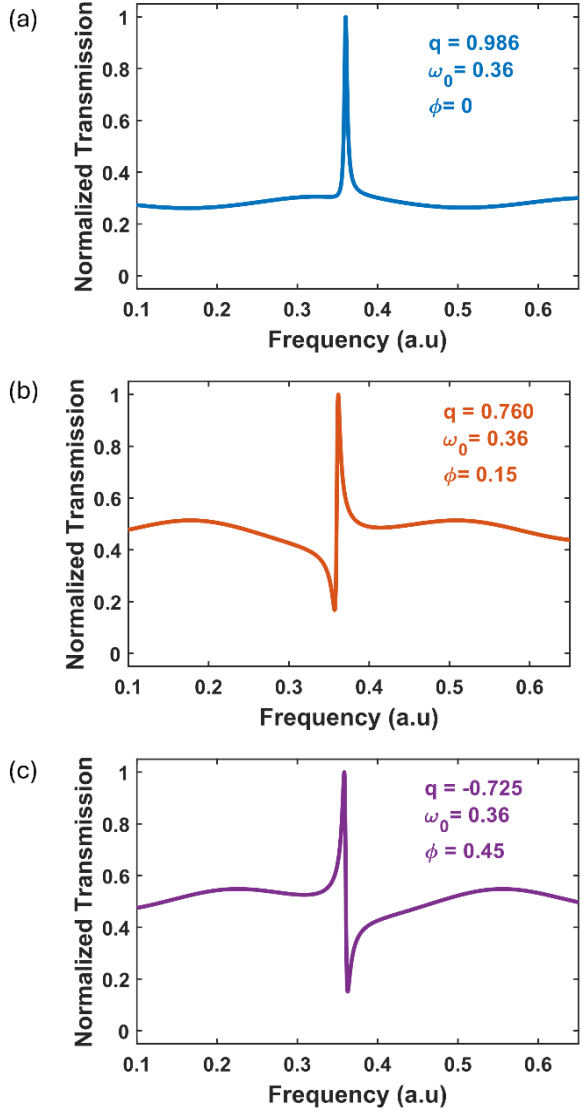

$$t_{FP} = \frac{(1-r^2)e^{-i\delta/2}}{r^2 e^{-i\delta} - 1} \quad (8)$$

where, $\delta/2$ denotes the phase shift accumulated as the waveguide mode propagates between the two partially reflecting elements. By plotting the transmission spectrum using Eq. (8), we observe an oscillatory response, as shown in Fig. S3d. Now, if we introduce an optical cavity between these reflective elements-as illustrated in the schematic of Fig. 3e. the system can be described using TCMT,

$$\frac{d\alpha}{dt} = -i(\omega - \omega_0)\alpha - \alpha\gamma_t + \sqrt{2\gamma_1}S_1^+ \quad (9)$$

$$S_1^- = r_c S_1^+ + \sqrt{2\gamma_1}\alpha \quad (10)$$

$$S_2^- = \sqrt{C_{11}}\sqrt{2\gamma_2}\alpha + \sqrt{C_{12}}t_c S_1^+ \quad (11)$$

Here, $r_c$ and $t_c$ denote the non-resonant reflection and transmission amplitudes, respectively. The coefficients $C_{11}$ and $C_{12}$ represent the relative contributions of the two interfering optical pathways. Using Equations (9) and (11), the total transmitted amplitude can be expressed as:

**Fig.S4.(a-c)** Tuning of Fano parameter due to different phases of the background oscillatory motion.

$$t(\omega) = \frac{|S_2^-|}{|S_1^+|} = \sqrt{C_{11}} \frac{2\sqrt{\gamma_1\gamma_2}}{i(\omega-\omega_0)+\gamma_t} + \sqrt{C_{12}}t_c$$

$$t(\omega) = \sqrt{C_{11}} \frac{2\sqrt{\gamma_1\gamma_2}}{i(\omega-\omega_0)+\gamma_t} + \sqrt{C_{12}} \frac{(1-r^2)e^{-i\delta/2}}{r^2 e^{-i\delta} - 1} \quad (12)$$

The transmission intensity as a function of frequency $\omega$ is plotted in fig.1d. The diagram in Figure S3e. includes a partially reflecting element, which introduces an oscillatory background. This background, when interacting with the discrete cavity resonance, gives rise to Fano interference, resulting in the combined spectrum. We have plotted eq. (12) which is shown on Fig.S3f.

Figure S4(a-c) presents an analytical model illustrating the variation of the Fano asymmetry parameter while keeping the cavity resonance frequency fixed. The phase of the oscillatory background is varied to mimic the effect of fiber-position-induced phase changes, reproducing the experimentally observed evolution of the Fano lineshape and confirming that the tuning of the Fano parameter arises from interference control rather than intrinsic cavity resonance shift.

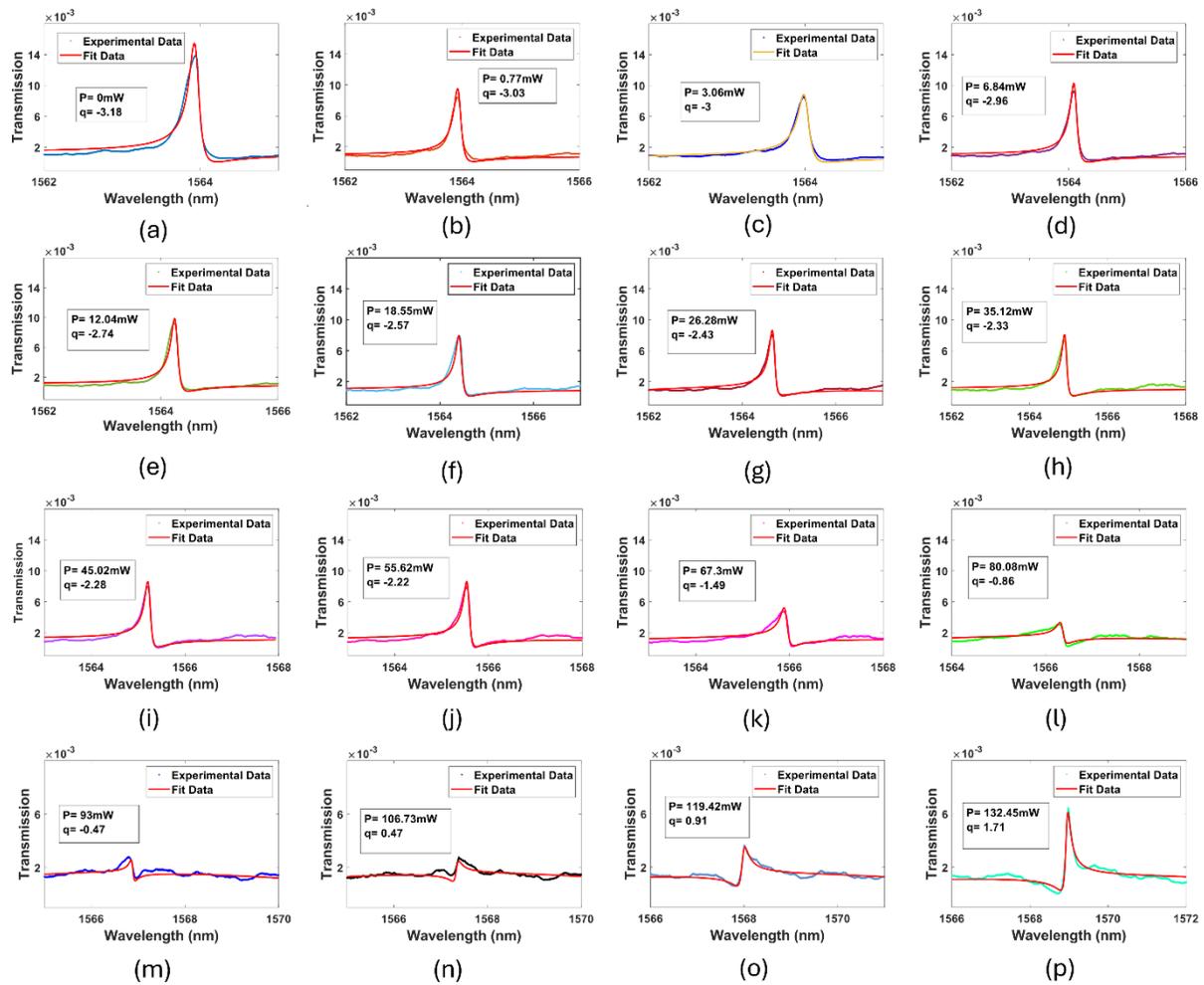

**Fig.S5.(a-p)** Tuning of Fano parameter due to different electrical heater power

Figure S5(a–p) shows the evolution of the Fano asymmetry parameter for all applied heater power levels. A total of 16 experimental transmission spectra is included, demonstrating continuous tuning of the Fano parameter from −3.18 to +1.71. The corresponding Fano fits are used to extract the parameter at each heating condition. Similar tuning behaviour was consistently observed across multiple fabricated devices, confirming the reproducibility of the interference-controlled response.